\documentclass[aps,prd,nofootinbib,showpacs,floatfix,superscriptaddress,preprintnumbers]{revtex4}
\usepackage{amsmath}
\usepackage{amssymb}
\usepackage{epsfig}
\usepackage{graphicx}
\usepackage{pstricks}
\usepackage{pst-coil}
\DeclareMathAlphabet{\mathsc}{OT1}{cmr}{m}{sc}
\newcommand {\ignore}[1]{}

\def\10{$SO(10)$}
\def\21{SU(2) $\otimes$ U(1) }

\def\422{$SU(4) \otimes SU(2) \otimes SU(2)$}
\def\321{SU(3) $\otimes$ SU(2) $\otimes$ U(1)}
\def\gsim{\raise0.3ex\hbox{$\;>$\kern-0.75em\raise-1.1ex\hbox{$\sim\;$}}}
\def\lsim{\raise0.3ex\hbox{$\;<$\kern-0.75em\raise-1.1ex\hbox{$\sim\;$}}}

\def\lsim{\raise0.3ex\hbox{$\;<$\kern-0.75em\raise-1.1ex\hbox{$\sim\;$}}}
\def\gsim{\raise0.3ex\hbox{$\;>$\kern-0.75em\raise-1.1ex\hbox{$\sim\;$}}}
\def\vev#1{\left\langle #1\right\rangle}
\def \znbb {0\nu\beta\beta}

\newcommand{\AddrAHEP}{%
  AHEP Group, Institut de F\'{\i}sica Corpuscular --
  C.S.I.C./Universitat de Val{\`e}ncia \\
  Edificio Institutos de Paterna, Apt 22085, E--46071 Valencia, Spain}

 \newcommand{\ba}{\begin{array}}
\newcommand{\ea}{\end{array}}
\relax

\def\321{$SU(3)\times SU(2)\times U(1)$}

\begin{document}


\renewcommand{\Huge}{\Large}
\renewcommand{\LARGE}{\Large}
\renewcommand{\Large}{\large}
\def \znbb {$0\nu\beta\beta$ }
\def \nbb {$\beta\beta_{0\nu}$ }
\title{Model for T2K indication with maximal $\theta_{23}$ and tri-maximal $\theta_{12}$}  \date{\today}
\author{S. Morisi} \email{morisi@ific.uv.es} \affiliation{\AddrAHEP}
\author{Ketan M. Patel} \email{kmpatel@prl.res.in} \affiliation{Physical Research Laboratory, Navarangpura, Ahmedabad-380 009, India.}
\author{E. Peinado} \email{epeinado@ific.uv.es} \affiliation{\AddrAHEP}


\date{\today}

\begin{abstract}
Recently T2K gives hint in favor of large reactor angle $\theta_{13}$. Most of the models, with
tri-bimaximal mixing at the leading order, can not reproduce such a large mixing angle since they
predict typically corrections for the reactor angle of the order $\theta_{13}\sim \lambda_C^2$,
where $\lambda_C\sim 0.2$. In this paper, we discuss the possibility to achieve large $\theta_{13}$
within the T2K region with maximal atmospheric mixing angle, $\sin^2\theta_{23}= 1/2$, and
trimaximal solar mixing angle, $\sin^2\theta_{12} = 1/3$, through the deviation from
the exact tri-bimaximal mixing. We derive the structure of neutrino mass matrix that leads to the
large $\theta_{13}$ leaving maximal $\theta_{23}$ and trimaximal $\theta_{12}$. It is shown that
such a structure of neutrino mass matrix can arise in a model with $S_4$ flavor symmetry.

\end{abstract}


\maketitle

\section{Introduction}
The T2K collaboration recently gives that the reactor angle is \cite{Abe:2011sj}
\begin{equation}\label{exp}
0.087 (0.100)\le \sin\theta_{13}\le 0.275 (0.306),
\end{equation}
with best fit value of $\sin \theta_{12}=0.17(0.19)$ for normal (inverted) hierarchy in the neutrino masses. Such a result must be taken as a hint since
it is only at 2$\sigma$. We however consider the implications of such an important indication in
this paper.

In the last decade, the tri-bimaximal (TB) mixing pattern introduced by Harrison {\it et al.} in
2002\,\cite{Harrison:2002er}
\begin{equation}
U_{\text{TB}}=
\left(
\begin{array}{ccc}
\sqrt{\frac{2}{3}}&\frac{1}{\sqrt{3}}& 0\\
-\frac{1}{\sqrt{6}} &\frac{1}{\sqrt{3}} & -\frac{1}{\sqrt{2}}\\
-\frac{1}{\sqrt{6}} &\frac{1}{\sqrt{3}} & \frac{1}{\sqrt{2}}\\
\end{array}
\right),
\end{equation}
has been used as a guide in neutrino physics for the flavor problem. However if the result of T2K
will be confirmed, it will have strong impact on this point of view. In fact, most of the models
predicting tri-bimaximal mixing at the leading order are compatible
only with small values of the reactor angle. In a generic model, the three mixing angles receive
corrections of the same order. Departure of the solar angle from the trimaximal values are at
most of $\mathcal{O}$($\lambda_C^2$) where
$\lambda_C\approx 0.2$. Therefore it is possible
to have a deviation of order $\lambda_C^2 \approx 0.04$\, only \cite{Altarelli:2010gt} for the
reactor angle which is about half of the lower bound in\,(\ref{exp}). This is only an estimation
and it must be considered individually case by case. We however expect that most of them are on the
border of validity if not excluded completely. Therefore it is important to search for the models
with large $\theta_{13}$ and with maximal atmospheric mixing angle and trimaximal\footnote{The word
``trimaximal'' is used for different meaning in previous studies \cite{trimaximal}. We mean here
$\sin^2\theta_{12} = 1/3$ by trimaximal solar angle.} solar angle. Recently, some works fitting T2K
result has been presented \cite{large13}. There are some models based on discrete flavor symmetries before T2K data which predicts large reactor mixing angle, for an incomplete list see reference \cite{Hirsch:2007kh}, and  for a classification of models with flavor symmetries classified by its predictions for reactor angle see \cite{Albright:2009cn}

In this paper, we study the possibility to obtain the large reactor angle with tri-bimaximal
values of the solar and atmospheric mixing angles. The lepton mixing matrix with such mixing
pattern was first proposed by King in \cite{King:2009qt} and called Tri-bimaximal-reactor (TBR)
mixing. Using such mixing matrix, we found the structure of the deviations in the neutrino mass
matrix from its TB texture which leads to TBR mixing. We then show that such a particular deviation
in neutrino mass matrix can arise in a model with $S_4$ flavor symmetry.

The paper is organized as follows. We discuss the conditions to obtain a mass matrix with maximal
atmospheric mixing angle, trimaximal solar mixing angle and a non-zero reactor mixing angle within
the T2K region in section \ref{LR}. In section \ref{model}, we present a model based on the group
of permutation of four objects, $S_4$ where the neutrino mass matrix with particular form discussed
in section \ref{LR} is obtained. Finally, we discuss the phenomenology of the model and conclude in
in section \ref{pheno}.

\section{Large Reactor Tri-Bimaximal mixing and neutrino mass matrix}
\label{LR}
In this section, we study the structure of the neutrino mass matrix (in the diagonal basis of the
charged leptons) that gives maximal atmospheric angle $\theta_{23}= \pi/4$, trimaximal
solar angle $\sin\overline{\theta}_{12}= 1/\sqrt{3}$ and an arbitrary reactor angle
$\theta_{13}=\lambda$. In the standard PDG \cite{Nakamura:2010zzi} parametrization,
the lepton mixing matrix with the above values of mixing angles is given by \cite{King:2009qt}
\begin{equation}\label{LRTB}
U_{\text{TBR}}=R_{23} (\frac{\pi}{4})\,R_{13}(\lambda) \,R_{12}(\overline{\theta}_{12})=
\left(
\begin{array}{ccc}
\sqrt{\frac{2}{3}}&\frac{1}{\sqrt{3}}& \lambda\\
-\frac{1}{\sqrt{6}}+\frac{\lambda}{\sqrt{3}} &\frac{1}{\sqrt{3}}+\frac{\lambda}{\sqrt{6}} &
-\frac{1}{\sqrt{2}}\\
-\frac{1}{\sqrt{6}}-\frac{\lambda}{\sqrt{3}} &\frac{1}{\sqrt{3}}-\frac{\lambda}{\sqrt{6}} &
\frac{1}{\sqrt{2}}\\
\end{array}
\right)+\mathcal{O}(\lambda^2).
\end{equation}
We do not consider the CP violation in the lepton sector assume that the above
parameters are real for simplicity. The neutrino mass matrix diagonalized by\,(\ref{LRTB}) is
given by
\begin{equation}\label{mLRTB}
m_\nu^{\text{TBR}}=U_{\text{TBR}}\cdot m_\nu^{\text{diag}} \cdot U_{\text{TBR}}^T= m_\nu^{TB}+\delta
m_\nu
\end{equation}
where $m_\nu^{\text{diag}}$ is a diagonal matrix with the neutrino mass eigenvalues, $m_{\nu_1}$,
$m_{\nu_2}$ and $m_{\nu_3}$. This leads to the following structure of the neutrino mass matrix
\begin{equation}\label{mntbm}
m_\nu^{\text{TB}}=
\left(
\begin{array}{ccc}
2y-x & x&x\\
x & y+z&y-z\\
x & y-z&y+z
\end{array}
\right),
\end{equation}
where $x=(m_2-m_1)/3$, $y=(m_1+2m_2)/6$ and $z=m_3/2$ and
\begin{equation}\label{deviation}
\delta m_\nu=\lambda
\left(
\begin{array}{ccc}
0 & \alpha_1& -\alpha_1\\
\alpha_1 & \beta_1&0\\
-\alpha_1 & 0& -\beta_1
\end{array}
\right)+
\lambda^2 \left(
\begin{array}{ccc}
\gamma & \alpha_2 & \alpha_2\\
\alpha_2 & \beta_2&-\beta_2\\
\alpha_2 & -\beta_2& \beta_2
\end{array}
\right)+
\sum_{n\ge 3} \lambda^n\left(
\begin{array}{ccc}
0 & \alpha_n& (-1)^n\alpha_n\\
\alpha_n & 0&0\\
(-1)^n\alpha_n & 0& 0
\end{array}
\right),
\end{equation}
with $\alpha_1=-(x-2y+2z)/\sqrt{2}$, $\beta_1=\sqrt{2}x$, $\alpha_2=-x/2$, $\beta_2=-(x-2y+z)/2$
and $\gamma=x-2y+2z$. Note that $\beta_2$ can be reabsorbed into the TB term $m_\nu^{\text{TB}}$.
The above form of neutrino mass matrix predicts maximal atmospheric mixing angle and trimaximal
solar mixing angle if all the terms with all powers of $\lambda$ are taken into account.
If one truncates the series in eq.\,(\ref{deviation}) at $n < 3$, the neutrino mass matrix then
implies
\begin{itemize}
\item (A) negligible deviations from maximality in the atmospheric mixing angle;
\item (B) small deviation from trimaximality in the solar mixing angle;
\item (C) prediction of $0\nu\beta\beta \propto \lambda^2$.
\end{itemize}
The prediction (C) is evident from eq.\,(\ref{deviation}) and we verify numerically (A) and (B) in
the section \ref{pheno}. We observe that the main structure of the deviation $\delta
m_\nu$ of order $\lambda$ in eq.\,(\ref{deviation}) is $\mu$-$\tau$ antisymmetric,
see\,\cite{Grimus:2006jz}\footnote{Note that the main structure of the deviation $\delta
m_\nu$ of order $\lambda$ in eq.\,(\ref{deviation}) is similar to the one found in the paper by T. Araki in Ref.\cite{large13} where (contrary with respect to us) the solar angle is not fixed to be the
trimaximal one. }. Therefore a possible flavor symmetry with neutrino mass matrix texture
(\ref{mLRTB}) must contain the group $S_2$ of the $\mu$-$\tau$ permutation and must be
compatible with tri-bimaximal in the unperturbed limit. One possible flavor symmetry with such
features is $S_4$ which contains $S_2$ as a subgroup and leads to tri-bimaximal
mixing\,\cite{Lam:2008sh}.

\section{The Model}
\label{model}
We assume $S_4$ (see appendix) flavor symmetry and extra abelian $Z_N$ symmetry in order to
separate the charged leptons from the neutrino sector as usual in models for TB mixing, see for
instance \cite{Altarelli:2010gt}. In order to simplify the model as much as possible and to render
more clear the main features of the model, we do not enter into the details of the particular
$Z_N$ symmetry required in this model. Our purpose is to show that the neutrino mass matrix
(\ref{mLRTB}) with the structure given by (\ref{mntbm}) and (\ref{deviation}) can be obtained from
symmetry principle. We assume that light neutrino masses arise from both type-I and type-II seesaw
and introduce only one right-handed neutrino. The matter content of our model is given in
table\,\ref{tab1}.
\begin{table}[h!]
\begin{tabular}{|c|c|c|c|c||c|c||c|c|}
\hline
 & $L$ & $l_R$&$\nu_R$  & $h$ & $\Delta_{}$ &  $\phi_{}$  & $\varphi_l$ & $\xi_l$ \\
\hline
$SU_L(2)$ & 2 & 1 & 1 & 2 & 3  & 1 &1&1\\
\hline
$S_4$ & $3_1$ & $3_1$ & $1_1$ & $1_1$ & $3_1$  & $3_1$ & $2$&$1_1$\\
\hline
\end{tabular}
\caption{Matter content of the model giving TB mixing at the leading order}\label{tab1}
\end{table}

In the scalar sector, we have one $SU_L(2)$ triplet $\Delta$ and one singlet $\phi$ in the
neutrino sector transforming both as $3_1$ of $S_4$. We have two electroweak singlets $\varphi_l$
and $\xi_l$ in the charged lepton sector, transforming as doublet and singlet of $S_4$
respectively. As it has been already mentioned, the two sectors can be separated by introducing an
abelian $Z_N$ symmetry under which $l^c$, $\varphi_l$ and $\xi_l$ are charged while the other
fields could be singlets of $Z_N$. The Yukawa interaction of the model is
\begin{eqnarray}
-\mathcal{L}_l   &=& \frac{1}{\Lambda} y_1 (\overline{L}l_R)_{1_1}h\xi_l + \frac{1}{\Lambda} y_2
(\overline{L}l_R)_2h\varphi_l+h.c.\\
-\mathcal{L}_\nu &=& y_a LL\Delta + \frac{y_b}{\Lambda} (\overline{L}
\phi)_{1_1}\tilde{h}\nu_R+\frac{1}{2}M\nu^c\nu^c+h.c.
\end{eqnarray}
where $\Lambda$ is an effective scale. We assume the following $S_4$ alignment in the vacuum
expectation values (vevs) of the scalar fields.
\begin{eqnarray}
\vev{\Delta^0}=v_\Delta (1,1,1)^T,\quad \vev{ \phi}=v_\phi (0,1,-1)^T,\quad \vev{\varphi} =
(v_1,v_2)^T,
\end{eqnarray}
where $v_1\ne v_2$. Using the product rules shown in appendix A, one can easily see that the
charged lepton mass matrix is diagonal and the lepton masses can be fitted in terms of three free
parameters $y_1$, $v_1$ and $v_2$, see \cite{Morisi:2011ge} for details.

The type-II seesaw gives a contribution to the neutrino mass matrix with zero diagonal entries
and equal off diagonal entries since it arises from the product of three $S_4$ triplets. Since
we introduced only one right-handed neutrino, Dirac neutrino mass matrix is a column
$m_D\sim (0,1,-1)^T$ and the light-neutrino mass matrix from seesaw relation is given by
\begin{equation}\label{mntbm2}
m_\nu^{\text{type-I}}= \frac{1}{M}m_D\,m_D^T\sim
\left(
\begin{array}{ccc}
0 & 0&0\\
0& 1&-1\\
0&-1&1
\end{array}
\right)
\end{equation}
Considering both the type-I and type-II contributions, we have the light neutrino mass
matrix which can be diagonalized by TB mixing matrix \cite{Dutta:2009bj}\footnote {In
\cite{Dutta:2009bj}, similar structure (\ref{mntbm2}) has been obtained through only type-II
seesaw and $S_4$ symmetry.}
\begin{equation}\label{mn0}
m_\nu^{\text{TB}} =
\left(
\begin{array}{ccc}
0&a&a\\
a&b&a-b\\
a&a-b&b
\end{array}
\right),
\end{equation}
The mass eigenvalues of the above matrix are $m_1=-a$, $m_2=2a$ and $m_3=-a+2b$.
Here $a= y_a v_\Delta $ and $b=y_b^2 v_h^2 v_\phi^2/(\Lambda^2 M)$ where $v_h=\vev{h^0}$. This neutrino mass matrix is compatible with the normal hierarchy only and predicts zero neutrinoless double beta decay $m_{ee}=0$.

In order to reproduce deviations like eq.\,(\ref{deviation}) in the neutrino mass matrix, we
introduce in the scalar sector one Higgs triplet $\Delta_d$ that transforms as a doublet
under $S_4$ and an electroweak singlet $\phi_d$ that transforms as a triplet $3_1$ under $S_4$.
With inclusion of these fields, the Yukawa interaction Lagrangian $\mathcal{L}_\nu$ contains also
the terms
\begin{equation}\label{lag2}
-\mathcal{L}_\nu\supset y_\beta LL\Delta_d+ \frac{y_\alpha}{\Lambda}
(\overline{L}\phi_d)_{1_1} \tilde{h}\nu_R+h.c.
\end{equation}
We assume that $\Delta_d$ and $\phi_d$ take vevs along the following directions
\begin{equation}\label{all2}
\vev{\Delta_d^0}=v_d (1,0)^T,\quad  \vev{\phi_d}=u_d(1,0,0)^T.
\end{equation}
Here we also assume that $y_{\alpha,\beta}\ll y_{a,b}$. This can be realized assuming that
$\Delta_d$ and $\phi_d$ are charged under some extra abelian symmetry like $Z_N$ or $U_{FN}(1)$.

After electroweak symmetry breaking and integrating out the right-handed neutrino, eq. (\ref{lag2}) gives the following contribution to the neutrino mass matrix
\begin{equation}\label{eff}
\frac{y_by_\alpha v_h^2}{\Lambda^2 M}(\nu \phi)_{1_1}(\nu \phi_{\text{d}})_{1_1}+
\frac{y_\alpha^2 v_h^2}{\Lambda^2 M}(\nu \phi_{\text{d}})_{1_1}(\nu \phi_{\text{d}})_{1_1}.
\end{equation}
The second term in eq.\,(\ref{eff}) is smaller with respect to the first since we have assumed $y_\alpha \ll y_b$. In particular assuming $y_b\sim 1$ and $y_\alpha\sim\lambda$ the first term is proportional to $\lambda$ and the second term is proportional to $\lambda^2$.
The extra contributions to the neutrino mass matrix from the type-I see-saw are as follows
\begin{equation}
\delta m_\nu^{\text{type-I}}\sim
c_1\lambda \left(
\begin{array}{ccc}
0 & 1& -1\\
1 & 0&0\\
-1 & 0& 0
\end{array}
\right)+
c_2\lambda^2  \left(
\begin{array}{ccc}
1 & 0& 0\\
0 & 0&0\\
0 & 0& 0
\end{array}
\right).
\end{equation}
where, $c_1$ and $c_2$ are coefficients of order ${\mathcal O}(1)$. From the extra type-II seesaw
term in eq. (\ref{lag2}) and using the vev alignments as in (\ref{all2}), the additional
contribution to the perturbed neutrino mass matrix will be proportional to
$\nu_1\nu_1-\nu_2\nu_2$, therefore the contribution to the neutrino mass matrix coming from
Type-II see-saw is
\begin{equation}
\delta m_\nu^{\text{type-II}}
\sim \left(
\begin{array}{ccc}
0 & 0& 0\\
0 & 1&0\\
0 & 0& -1
\end{array}
\right).
\end{equation}
Putting all these results together, the structure of the deviation in neutrino mass matrix
can be written is
\begin{equation}\label{dmn}
\delta m_\nu=
\left(
\begin{array}{ccc}
\gamma' & \alpha'& -\alpha'\\
\alpha' & \beta'&0\\
-\alpha' & 0& -\beta'
\end{array}
\right),
\end{equation}
where $\alpha'=y_by_\alpha v_h^2 v_\phi u_d/(\Lambda^2 M)$, $\beta'=y_\beta v_d$,
$\gamma'=y_\alpha^2 v_h^2 v_\phi u_d/(\Lambda^2 M)$.

The deviation obtained in our model  equal to the neutrino mass deviation in eq.\,(\ref{deviation})
truncated at $\lambda^2$ with $\alpha_2=0$. Such a difference does not modify significantly the
prediction of maximal atmospheric angle and trimaximal solar angle. In the next section, we study
the phenomenological implication of our neutrino mass texture.

\section{phenomenology}
\label{pheno}
Combining eq. (\ref{mn0}) and eq. (\ref{dmn}), the resulting neutrino mass matrix in our model is
\begin{equation}\label{last}
m_\nu =
\left(
\begin{array}{ccc}
\gamma'&a+\alpha'&a-\alpha'\\
a+\alpha'&b+\beta'&a-b\\
a-\alpha'&a-b&b-\beta'
\end{array}
\right).
\end{equation}
As mentioned earlier, we assume $y_a, y_b\sim {\mathcal O}(1)$ and $y_\alpha, y_\beta \sim
{\mathcal O}(\lambda)$ which implies the hierarchies in the elements $a,b \gg \alpha', \beta' \gg
\gamma'$ in the above structure. Since the $(m_\nu)_{11}$ entry is  $\gamma \sim
\mathcal{O}(\lambda^2)$, we have neutrino less double beta decay rate $m_{ee}\propto \lambda^2$ and
for small values of $\lambda$ as in the unperturbed case only the normal neutrino mass hierarchy
can be fitted. The neutrino mass matrix\,(\ref{last}) obtained from the model is equivalent
to the matrix\,(\ref{mLRTB}) up to the correction of $\mathcal{O}$($\lambda^2$), the neutrino mass
matrix\,(\ref{last}) is diagonalized by the mixing matrix\,(\ref{LRTB}) up to
$\mathcal{O}$($\lambda^2$) corrections. Note that the neutrino mass matrix\,(\ref{last}) obtained
in the model has 5 free parameters while the derived structure in eq.\,(\ref{LRTB}) has 4 real
parameters ($x,y,z$ and $\lambda$). We fix free parameters of eq.\,(\ref{last}) in terms of
the parameters of eq.\,(\ref{LRTB}) by comparing both the structures at each order of $\lambda$.
Comparing the leading order expressions of the neutrino mass matrix in eq.\,(\ref{mn0}) and
eq.\,(\ref{mntbm}), we restrict our parameters to be
\begin{equation}\label{relat0}
x=2y;~a=2y;~b=y+z.
\end{equation}
Further comparing the higher order terms in $\lambda$, we obtain the relations
\begin{equation}\label{relat0}
\alpha' = -\sqrt{2}z \lambda;~\beta' = 2\sqrt{2}y\lambda;~\gamma'= 2z\lambda^2.
\end{equation}

\begin{figure}[h!]
\includegraphics[width=8cm]{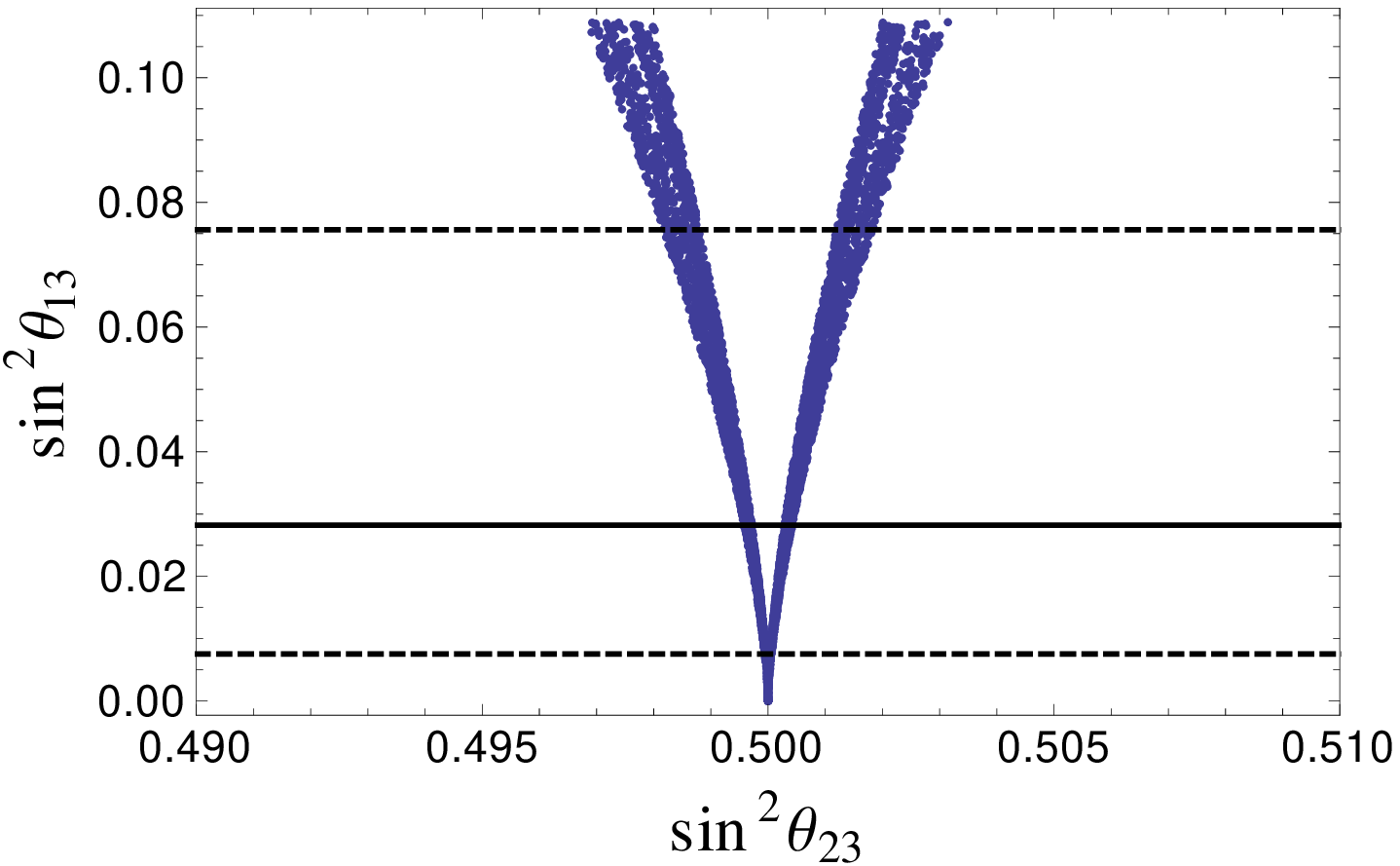}
\hspace{0.5cm}
\includegraphics[width=8cm]{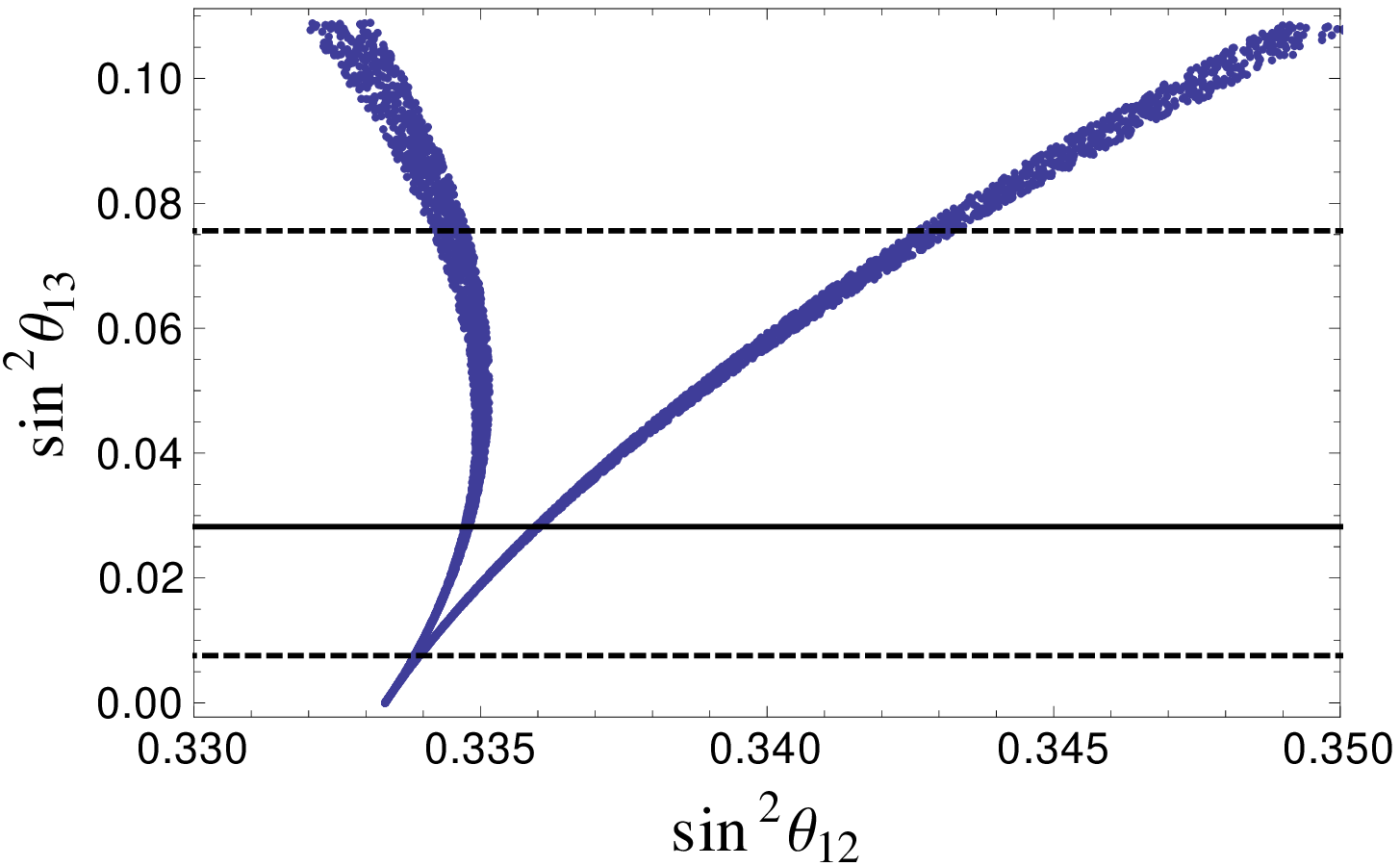}
\caption{The figure in the left side shows the allowed region for
$\theta_{13}~\mbox{vs.}~\theta_{23}$. The figure in the right side shows the allowed region for
$\theta_{13}~\mbox{vs.}~\theta_{12}$. The horizontal continuous (dashed) lines represent the best
fit value (2$\sigma$ deviations) of T2K for the reactor neutrino mixing angle.}
\label{figt}
\end{figure}

Note that this is not the most general case for our model, nevertheless we want to point out that
even in the case of bimaximal atmospheric mixing angle and trimaximal solar mixing angle it is
possible to obtain large reactor mixing angle. Also, we expect the negligible deviations from the
tri-bimaximal values for the solar and atmospheric mixing angles due to the fact that our model
predicts $\alpha_2=0$ if compared with eq.\,(\ref{mLRTB}) and it generates the terms only up to the
$\mathcal{O}$($\lambda^2$). We analyze such deviations by randomly varying $y$, $z$ and $\lambda$
with the constraints that the square mass differences and the mixing angles are in the observed
(3$\sigma$) range of validity \cite{Schwetz:2011qt}. The results of our analysis are shown in
figure \ref{figt}. It is evident from figure \ref{figt} that for restricted parameter space as
specified earlier, model allows large $\theta_{13}$ with negligible deviations in atmospheric and
solar mixing angles from there bimaximal and trimaximal values respectively. We also check the
predictions for the neutrinoless double beta decay rates in our model for restricted parameter
space specified above and find that the region for 4.5 meV $<m_{\nu1}<$ 5.8 meV and 0.5 meV $<
|m_{\beta \beta}|<$ 3.5 meV is allowed for the values of $\theta_{13}$ in the 2$\sigma$
limits indicated by T2K.

\vskip10.mm
In summary, we found the structure for the deviation in the neutrino mass matrix from the
well known TB pattern in such a way that the lepton mixing matrix has large
atmospheric mixing angle and trimaximal solar mixing angle with an arbitrary large reactor
angle. The deviation must be approximately $\mu$-$\tau$ antisymmetric. This fact suggests us that
the flavor symmetry could be some permutation symmetry containing $S_2$ ($\mu$-$\tau$ exchange)
subgroup. $S_3$ is too small since it does not give the TB mixing. The smallest permutation group
with this property is $S_4$. We provide a candidate model based on $S_4$ where in the unperturbed
limit the neutrino mass matrix is TB. Then assuming extra scalar fields we show the possibility to
generate deviations from the TB that give a large $\theta_{13}$ in agreement with T2K result,
maximal atmospheric mixing angle and trimaximal solar mixing angle in good agreement with neutrino
data.

\section{Acknowledgments}

This work was supported by the Spanish MICINN under grants FPA2008-00319/FPA and MULTIDARK
CSD2009-00064 (Consolider-Ingenio 2010 Programme), by Prometeo/2009/091 (Generalitat Valenciana),
by the EU Network grant UNILHC PITN-GA-2009-237920. S. M. is supported by a Juan de la Cierva
contract. E. P. is supported by CONACyT (Mexico). K. M. P. is grateful to IFIC, Universitat de
Val{\`e}ncia for hospitality and support.

\begin{appendix}

\section{$S_4$ product rules}
In the basis where the generator of $S_4$ are real, the products of $\mu \times \mu$ (see \cite{Lam:2008sh}):\\
\begin{center}\parbox{2.5in}{\begin{center}
 for  $2$
\begin{eqnarray}\nonumber
&a_1 a^{\prime}_1 + a_2 a^{\prime}_2 \sim 1_1,&\\ \nonumber
&-a_1 a^{\prime}_2 + a_2 a^{\prime}_1 \sim 1_2,&\\ \nonumber
&\left( \begin{array}{c} a_1 a^{\prime}_2 + a_2 a^{\prime}_1 \\ a_1 a^{\prime}_1 - a_2 a^{\prime}_2 \end{array} \right) \sim 2,&
\end{eqnarray}
\end{center}
}\end{center}
\parbox{3in}{\begin{center}
for $3_1$
\begin{eqnarray}\nonumber
&\sum \limits _{j=1} ^{3} b_j b^{\prime}_j  \sim 1_1,&\\ \nonumber
&\left( \begin{array}{c} \frac{1}{\sqrt{2}} (b_2 b^{\prime}
_2 - b_3 b^{\prime}_3) \\ \frac{1}{\sqrt{6}} (-2 b_1 b^{\prime}_1 + b_2 b^{\prime}_2 + b_3 b^{\prime}_3) \end{array} \right) \sim 2,& \\ \nonumber
&\left( \begin{array}{c} b_2 b^{\prime}_3 + b_3 b^{\prime}_2 \\ b_1 b^{\prime}_3 + b_3 b^{\prime}_1\\ b_1
    b^{\prime}_2 + b_2 b^{\prime}_1 \end{array} \right) \sim
3_1 \; , \;\; \left(
  \begin{array}{c} b_3 b^{\prime}_2 - b_2 b^{\prime}_3 \\ b_1 b^{\prime}_3 - b_3 b^{\prime}_1 \\ b_2 b^{\prime}_1 -
  b_1 b^{\prime}_2 \end{array} \right) \sim 3_2,&
\end{eqnarray}
\end{center}}
\parbox{3in}{\begin{center}
for $3_2$
\begin{eqnarray} \nonumber
&\sum \limits _{j=1} ^{3} c_j c^{\prime}_j \sim 1_1,&\\ \nonumber
&\left( \begin{array}{c} \frac{1}{\sqrt{2}} (c_2 c^{\prime}
_2 - c_3 c^{\prime}_3) \\ \frac{1}{\sqrt{6}} (-2 c_1 c^{\prime}_1 + c_2 c^{\prime}_2 + c_3 c^{\prime}_3) \end{array} \right) \sim 2,& \\ \nonumber
&\left( \begin{array}{c} c_2 c^{\prime}_3 + c_3 c^{\prime}_2 \\ c_1 c^{\prime}_3 + c_3 c^{\prime}_1\\ c_1
    c^{\prime}_2 + c_2 c^{\prime}_1 \end{array} \right) \sim 3_1\; , \;\;\left(
  \begin{array}{c} c_3 c^{\prime}_2 - c_2 c^{\prime}_3 \\ c_1 c^{\prime}_3 - c_3 c^{\prime}_1 \\ c_2 c^{\prime}_1 -
  c_1 c^{\prime}_2 \end{array} \right) \sim 3_2& .
\end{eqnarray}
\end{center}}\\
For $2 \times 3_1$:
\parbox{2.5in}{
\begin{eqnarray} \nonumber
\left( \begin{array}{c} a_2 b_1 \\ -\frac{1}{2}(\sqrt{3} a_1 b_2 + a_2
    b_2)\\  \frac{1}{2} (\sqrt{3} a_1 b_3 - a_2 b_3) \end{array}
\right) \sim 3_1\\ \nonumber
\left( \begin{array}{c} a_1 b_1 \\ \frac{1}{2}(\sqrt{3} a_2 b_2 - a_1
    b_2)\\  -\frac{1}{2} (\sqrt{3} a_2 b_3 + a_1 b_3) \end{array}
\right) \sim 3_2
\end{eqnarray}}
\parbox{1in}{and for  $2 \times 3_2$}
\parbox{2in}{
\begin{eqnarray} \nonumber
\left( \begin{array}{c} a_1 c_1 \\ \frac{1}{2}(\sqrt{3} a_2 c_2 - a_1
    c_2)\\  -\frac{1}{2} (\sqrt{3} a_2 c_3 + a_1 c_3) \end{array}
\right) \sim 3_1\\ \nonumber
\left( \begin{array}{c} a_2 c_1 \\ -\frac{1}{2}(\sqrt{3} a_1 c_2 + a_2
    c_2)\\  \frac{1}{2} (\sqrt{3} a_1 c_3 - a_2 c_3) \end{array}
\right) \sim 3_2 .
\end{eqnarray}}
\begin{center}
\hspace{-2.8in}
For $3_1\times 3_2$
\begin{eqnarray} \nonumber
& \sum \limits _{j=1} ^{3} b_j c_j \sim 1_2&\\ \nonumber
&\left( \begin{array}{c}  \frac{1}{\sqrt{6}} (2 b_1 c_1 - b_2 c_2 - b_3
    c_3) \\ \frac{1}{\sqrt{2}} (b_2 c
_2 - b_3 c_3) \end{array} \right) \sim 2& \\ \nonumber
&\left( \begin{array}{c} b_3 c_2 - b_2 c_3 \\ b_1 c_3 - b_3 c_1 \\ b_2 c_1 -
  b_1 c_2 \end{array} \right) \sim 3_1\; , \;\;\left(
  \begin{array}{c} b_2 c_3 + b_3 c_2 \\ b_1 c_3 + b_3 c_1\\ b_1
    c_2 + b_2 c_1 \end{array} \right) \sim 3_2& .
\end{eqnarray}
\end{center}

\end{appendix}

\end{document}